\documentclass[conference]{IEEEtran}
\IEEEoverridecommandlockouts
\usepackage{cite}
\usepackage{amsmath,amssymb,amsfonts}
\usepackage{algorithmic}
\usepackage{graphicx}
\usepackage{cuted}
\usepackage{stfloats}
\usepackage{epstopdf}
\usepackage{textcomp}
\usepackage{xcolor}
\usepackage{textgreek}
\usepackage{array}
\usepackage{multirow}
\usepackage{lmodern} 
\usepackage{bold-extra}  
\def\BibTeX{{\rm B\kern-.05em{\sc i\kern-.025em b}\kern-.08em
    T\kern-.1667em\lower.7ex\hbox{E}\kern-.125emX}}

\begin{document}
\title{An Area-Efficient 20-100-GHz \\ Phase-Invariant Switch-Type Attenuator \\ Achieving 0.1-dB Tuning Step in 65-nm CMOS }
\author{\IEEEauthorblockN{ Qingbin Li, and Jian Pang }
\IEEEauthorblockA{State Key Laboratory of Radio Frequency Heterogeneous Integration, Shanghai Jiao Tong University \\
No. 800, Dongchuan Road, Shanghai, 200240, China\\
pangjian@sjtu.edu.cn}
}
\maketitle

Accepted paper at IEEE UCMMT 2025\\
\begin{abstract}
This paper presents a switch-type attenuator working from 20 to 100 GHz. The attenuator adopts a capacitive compensation technique to reduce phase error. The small resistors in this work are implemented with metal lines to reduce the intrinsic parasitic capacitance, which helps minimize the amplitude and phase errors over a wide frequency range. Moreover, the utilization of metal lines also reduces the chip area. In addition, a continuous tuning attenuation unit is employed to improve the overall attenuation accuracy of the attenuator. The passive attenuator is designed and fabricated in a standard 65nm CMOS. The measurement results reveal a relative attenuation range of 7.5 dB with a continuous tuning step within 20-100 GHz. The insertion loss is 1.6-3.8 dB within the operation band, while the return losses of all states are better than 11.5 dB. The RMS amplitude and phase errors are below 0.15 dB and $\mathbf{1.6^{\circ}}$, respectively.
\end{abstract}

\section{introduction}
Attenuators are extensively employed in various RF and millimeter-wave (mm-wave) applications, including wireless communication and radar systems\cite{ref14,ref15,ref13}. In phased array systems, amplitude control serves multiple functions, such as calibrating signals across different array channels and implementing beam tapering to suppress sidelobes\cite{ref12}. Generally, high tuning accuracy, low tuning error and compact area are required to be realized by such attenuators.

Several types of digital step attenuators have been widely researched. Switch-type attenuators are commonly used for their broad attenuation range and minimal phase variation\cite{ref1,ref3,ref8}. They can operate over a wide band but suffer from high insertion loss due to series transistors. Additionally, the parasitic difference between the switch's on and off states introduces considerable phase variation. Distributed attenuators have low insertion loss (IL) due to the absence of series transistors \cite{ref2}. However, this topology requires a large chip area due to the quarter-wavelength transmission lines. Compared with the above various attenuators, VGAs can provide additional power gain\cite{ref11}. However, the VGAs require additional DC power consumption and exhibit limited bandwidth.

This paper introduces an attenuator working in wideband (20-100GHz). The capacitive compensation is introduced to reduce the phase error of the attenuator. The small resistors implemented by metal lines of $\mathrm{M_8}$ layer in the 2-dB attenuation unit reduce the inner parasitic capacitance, which contributes to reducing the amplitude and phase error over a wide frequency range. It also reduces the chip area. In addition, a continuous tuning unit is designed to improve the attenuation accuracy further. The isolated NMOS also reduce the IL. Transmission lines (TLs) between the attenuation units can improve the impedance-matching characteristic. In measurement, the proposed attenuator achieves a total attenuation of 7.5 dB within a bandwidth of 20 to 100 GHz. The IL of the attenuator is only 1.6-3.8 dB over the operation band. And the RMS gain error and RMS phase error of the attenuator are less than 0.15 dB and $1.6^\circ$, respectively.
\begin{figure}[tbp]
    \centering
    \includegraphics[width=0.5\textwidth]{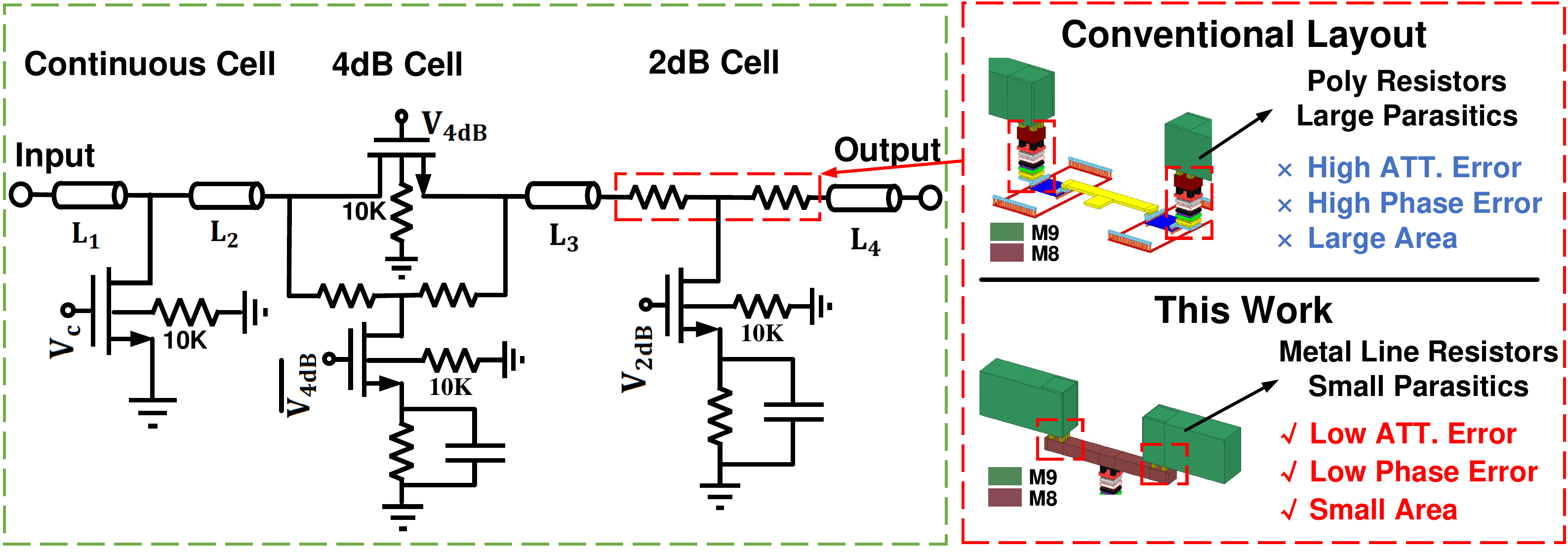}
    \caption{Schematic of the proposed attenuator and the small resistors implementation.}
    \label{core schematic}
\end{figure}

\section{analysis and design}
Fig. \ref{core schematic} shows the structure of the proposed attenuator. The 4-dB attenuation unit adopts T-type configuration, and 2-dB and continuous attenuation units adopt simplified T-type configurations. Series inter-stage impedance-matching TLs are inserted between attenuation units.

\subsection{T-type Attenuation Unit With Capacitive Compensation}  
\begin{figure}[tbp]
    \centering
    \includegraphics[width=0.48\textwidth]{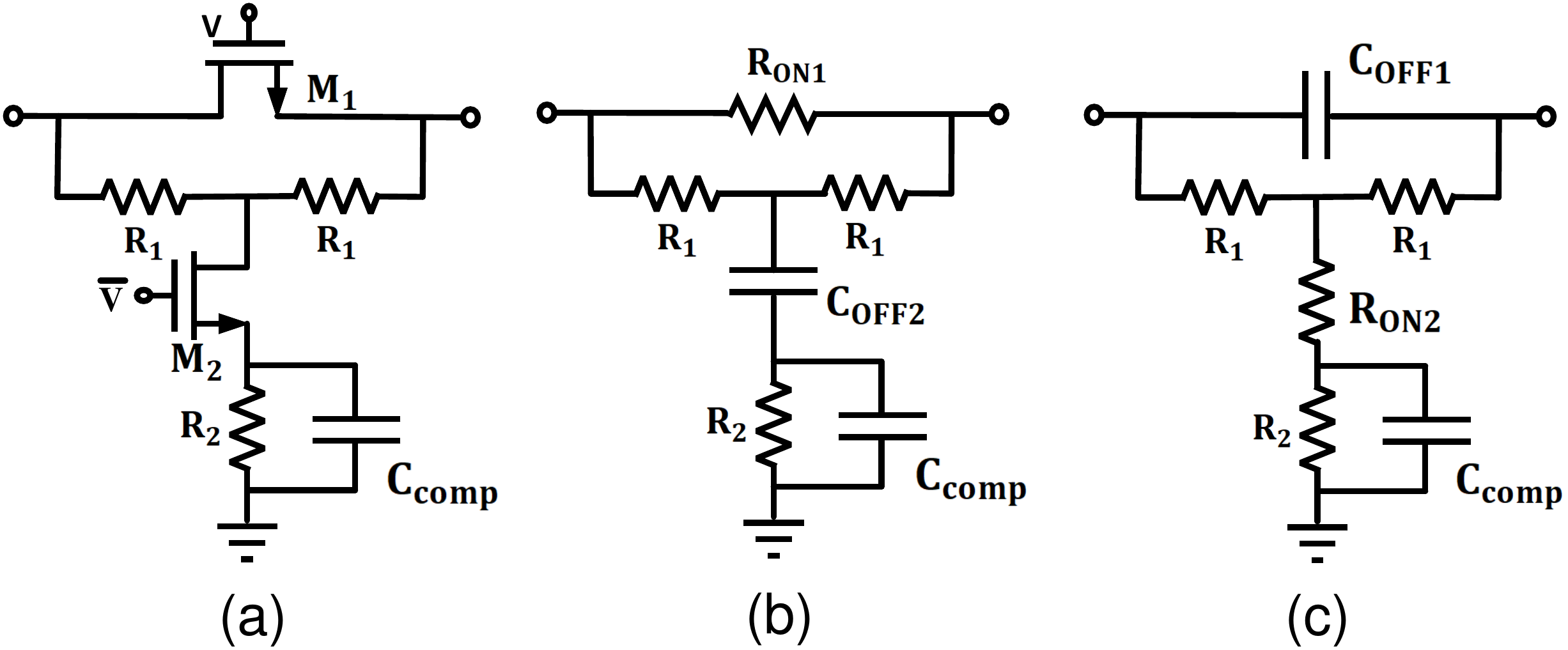}
    \caption{(a) Schematic of the T-type attenuator with capacitive compensation and its equivalent circuits, (b) reference state and (c) attenuation state.}
    \label{t-type}
\end{figure}

In this paper, a T-type attenuation unit with capacitor-paralleled compensation is introduced to compensate 
\begin{figure*}
\vspace*{-1.3cm}
  \begin{equation}
   S_{21} \simeq \frac{
  2Z_0(R_2 + R_{ON2}) + 2 j\omega C_{comp} Z_0 R_{ON2} R_2 }
  {(1 + j\omega C_{comp} R_2) \left[ 2R_{ON2} (Z_0 + R_1) + (Z_0 + R_1)^2 \right] + 2 R_2(Z_0 + R_1)}
  \end{equation}
  \begin{equation}
  \angle S_{21} \simeq \frac{\omega C_{comp} R_{ON2} R_2} {R_2 + R_{ON2}} -
  \frac{\omega C_{comp} (2R_{ON2} + R_1 + Z_{0}) R_2}{Z_0 + R_1 + 2R_{ON2} + 2R_2}
  \end{equation}
\hrulefill
\vspace*{-0.5cm}  
\end{figure*}
the phase error between the reference state and the attenuation state. Fig. \ref{t-type} shows the equivalent circuits of the proposed T-type attenuation unit with capacitive compensation technique. The capacitor of $\mathrm{C_{comp}}$ and the resistor of $\mathrm{R_2}$ are placed in parallel on the shunt branch of the T-type network. Using S-parameter analysis, the transmission coefficient $\mathrm{S_{21}}$ and phase response of the attenuation state can be expressed as (1) and (2)\cite{ref1}, shown at the top of the page. In the attenuation state, the capacitive compensation structure behaves as a low-pass network, introducing phase lag in the signal, which in turn reduces the phase difference with the reference state.

\subsection{2-dB Attenuation Unit}
When the required attenuation for a given unit is as low as 2 dB, the resistance of $\mathrm{R_1}$ shown in Fig. \ref{t-type}(a) becomes vary small. Therefore, the series transistor $\mathrm{M_1}$ shown in Fig. \ref{t-type} can be removed to achieve a wider operating bandwidth with virtually no increase in IL.
And in this work, the small resistor is realized using metal lines, as shown in Fig. \ref{core schematic}. By carefully optimizing the length and width of the metal lines, the attenuation unit achieves an accurate target attenuation. Utilizing the $\mathrm{M8}$ layer of metal lines reduces the parasitic capacitance between the poly resistors and the signal lines, thereby minimizing attenuation and phase variations over a wide frequency band (20-100 GHz). Additionally, the use of metal lines reduces the chip area. 

\subsection{Continuous Tuning Unit}
\begin{figure}
    \centering
    \includegraphics[width=0.48\textwidth]{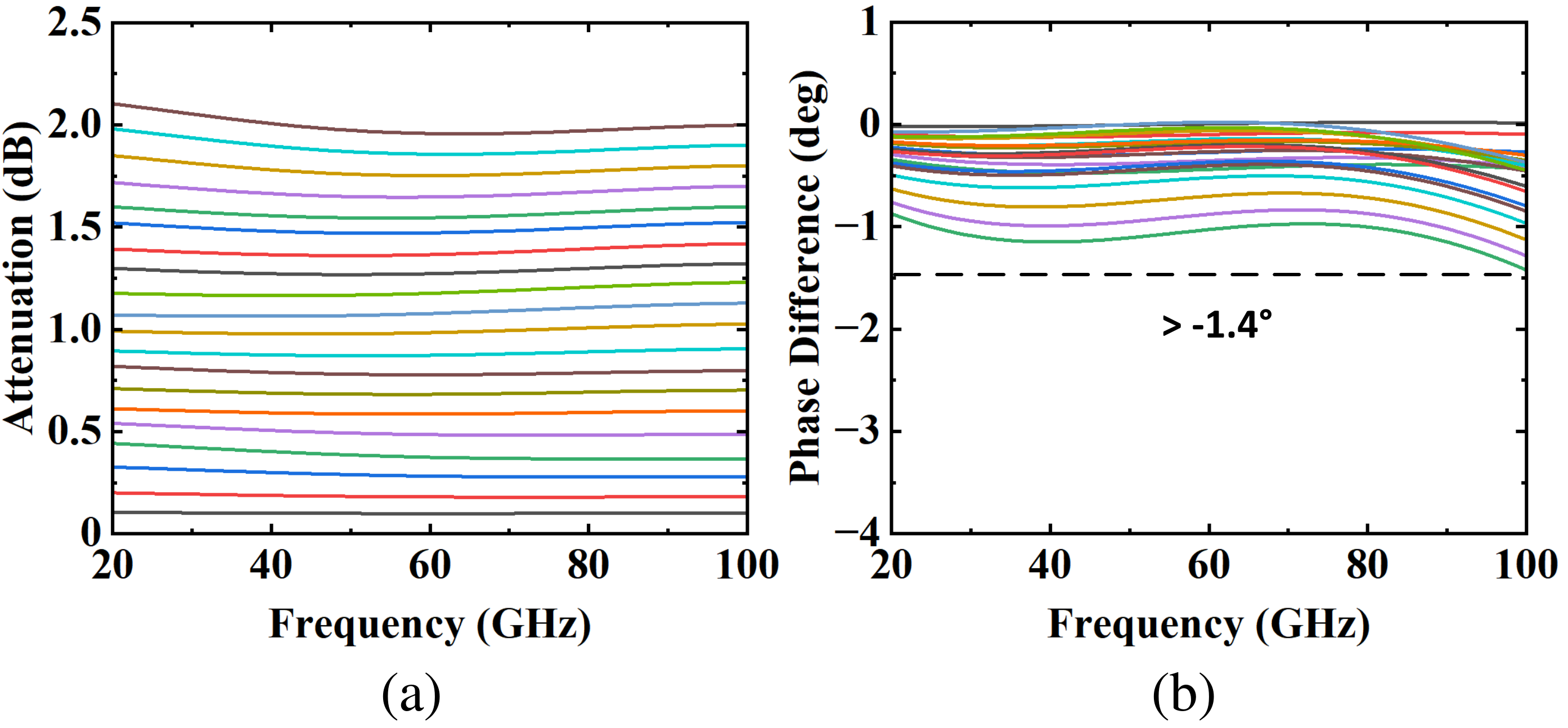}
    \caption{(a) Simulated attenuations of the continuous tuning unit (b) phase differences for all states. }
    \label{continuous}
\end{figure}

The continuous attenuation cell also adopts a simplified T-type topology, in which the resistor $\mathrm{R_1}$ is removed to further reduce the IL and improve return loss and linearity. By varying the control voltage of the shunt transistors, the attenuation cell achieves precise attenuation from 0 to 2 dB with a step of 0.1 dB. The simulated attenuation results and phase errors of all states are shown in Fig. \ref{continuous}. The use of this continuous attenuation cell enables more accurate attenuation for the proposed attenuator.

The proposed attenuator also utilizes isolated NMOS to decrease the IL caused by internal parasitic capacitance, which is especially important in cascaded structures \cite{ref10}. In addition, isolated NMOS contributes to a flatter gain characteristic by reducing the variation of the IL and the relative attenuation level becomes more consistent across the entire frequency band. 

Since the attenuator units exhibit capacitive impedance due to the parasitic capacitance, the matching performance will deteriorate at high frequencies. By inserting series TLs between the attenuation units, the matching performance can be improved at high frequencies and a great match can be achieved within the operation band. 

\section{Measurement Results}
\begin{figure}[tbp]
\centerline{\includegraphics[width=0.3\textwidth]{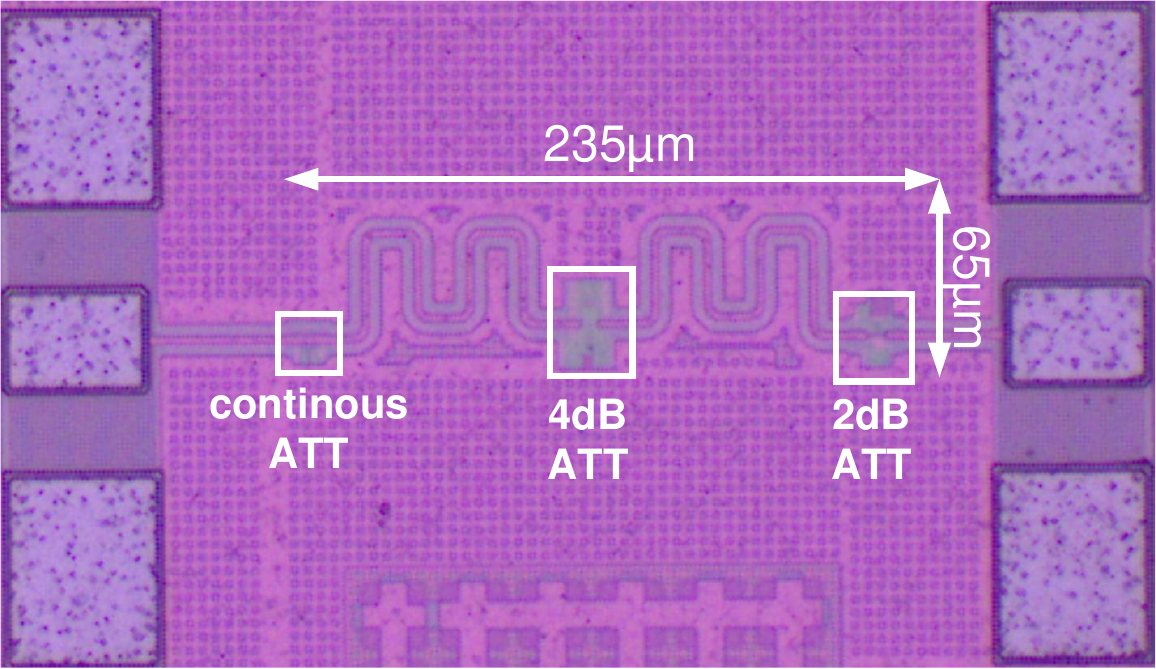}}
\caption{Chip photograph of the attenuator.}
\label{chip photograph}
\end{figure}

The proposed attenuator is fabricated in a 65nm CMOS process and the chip micrograph is shown in Fig. \ref{chip photograph}. The core chip area is only $0.015\; \mathrm{mm^2}$.

\begin{figure}[tbp]
\centerline{\includegraphics[width=0.5\textwidth]{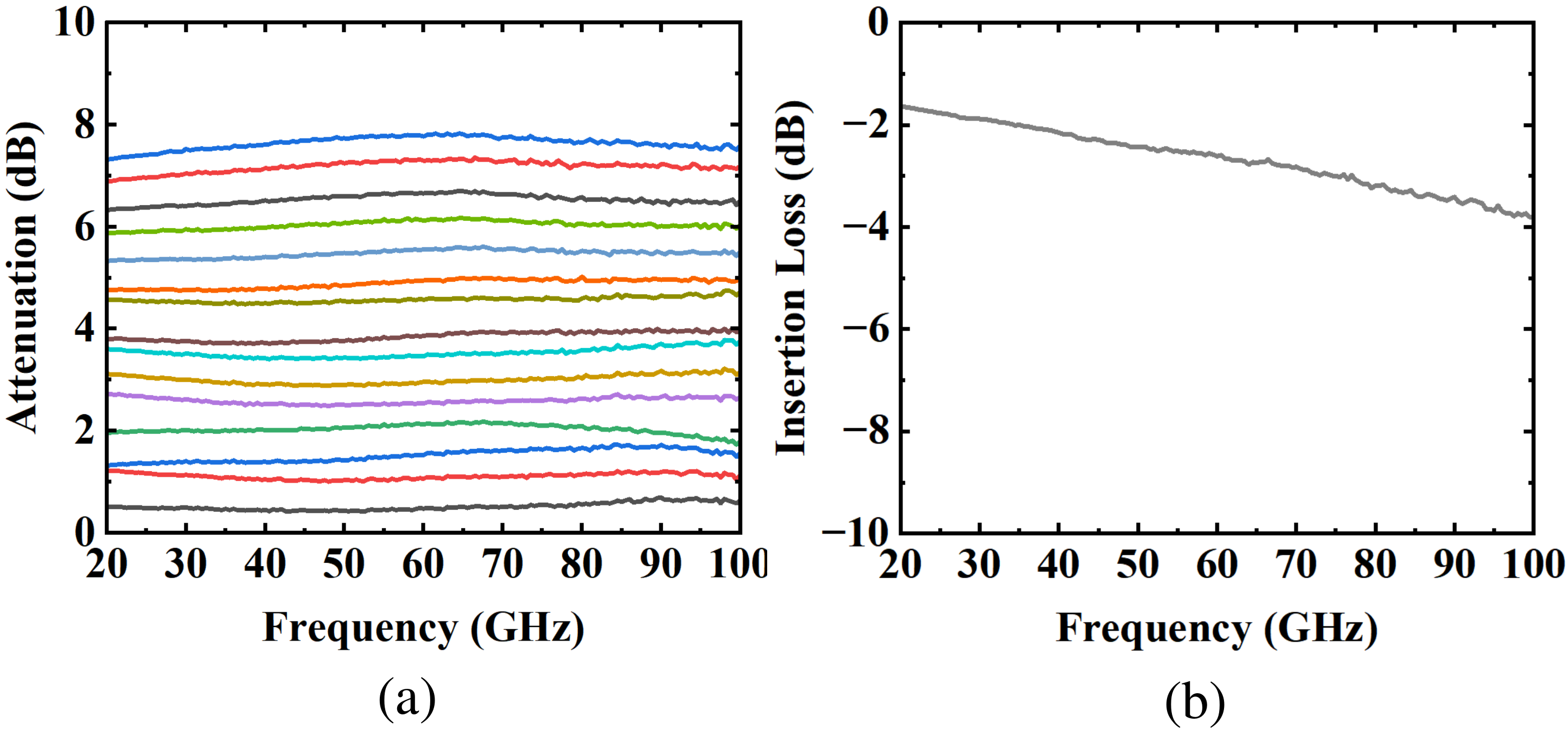}}
\caption{Measured (a) attenuations and (b) insertion loss.}
\label{ATT_IL}
\end{figure}

The S-parameter measurement is performed by using the Keysight N5227A network analyzer. The measured attenuation results over frequency for all states from 0 to 7.5 dB are shown in Fig. \ref{ATT_IL}{a}. Fig. \ref{ATT_IL}{b} shows that as the frequency increases, the parasitic effect results in more IL. The IL ranges from 1.6 to 3.8 dB across 20-100 GHz.

\begin{figure}[tbp]
\centerline{\includegraphics[width=0.5\textwidth]{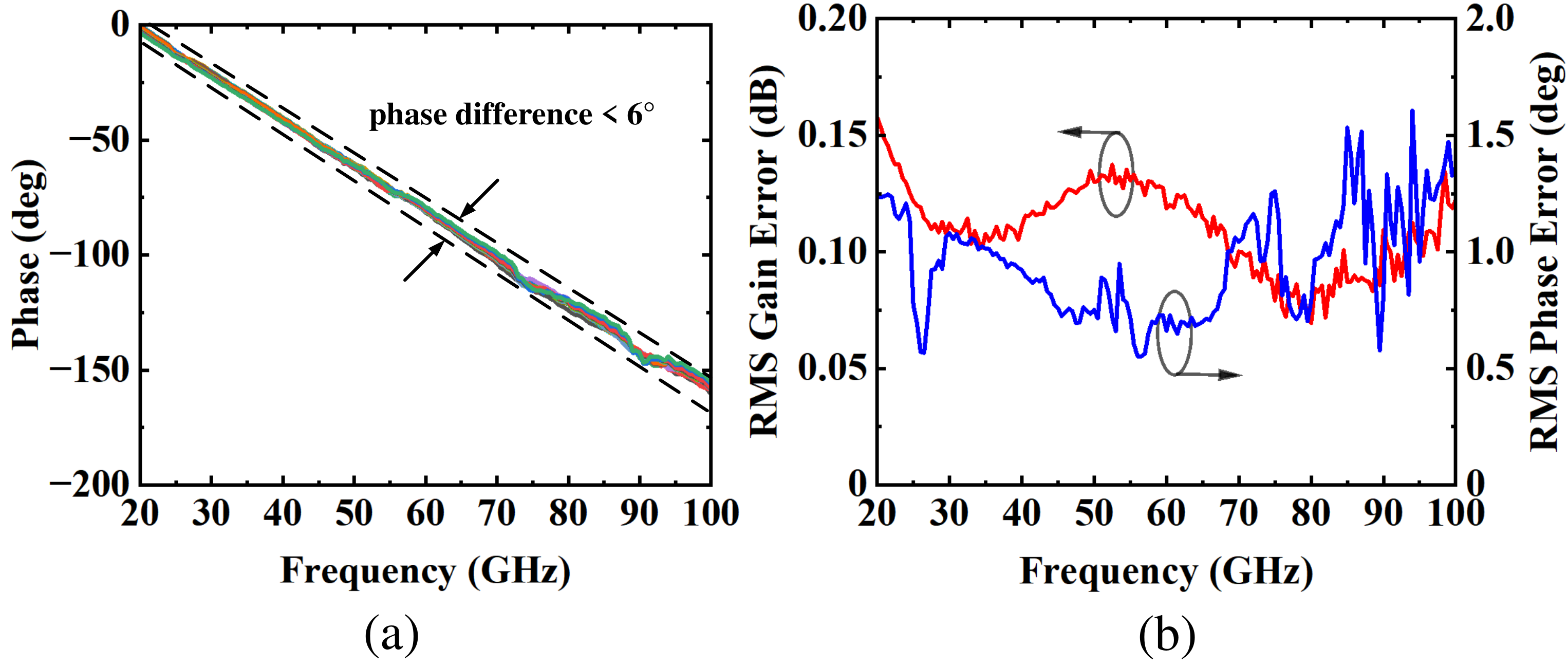}}
\caption{Measured (a) 16-state phases and (b) RMS gain/phase error.}
\label{Phase_RMS}
\end{figure}

\begin{figure}[tbp]
\centerline{\includegraphics[width=0.5\textwidth]{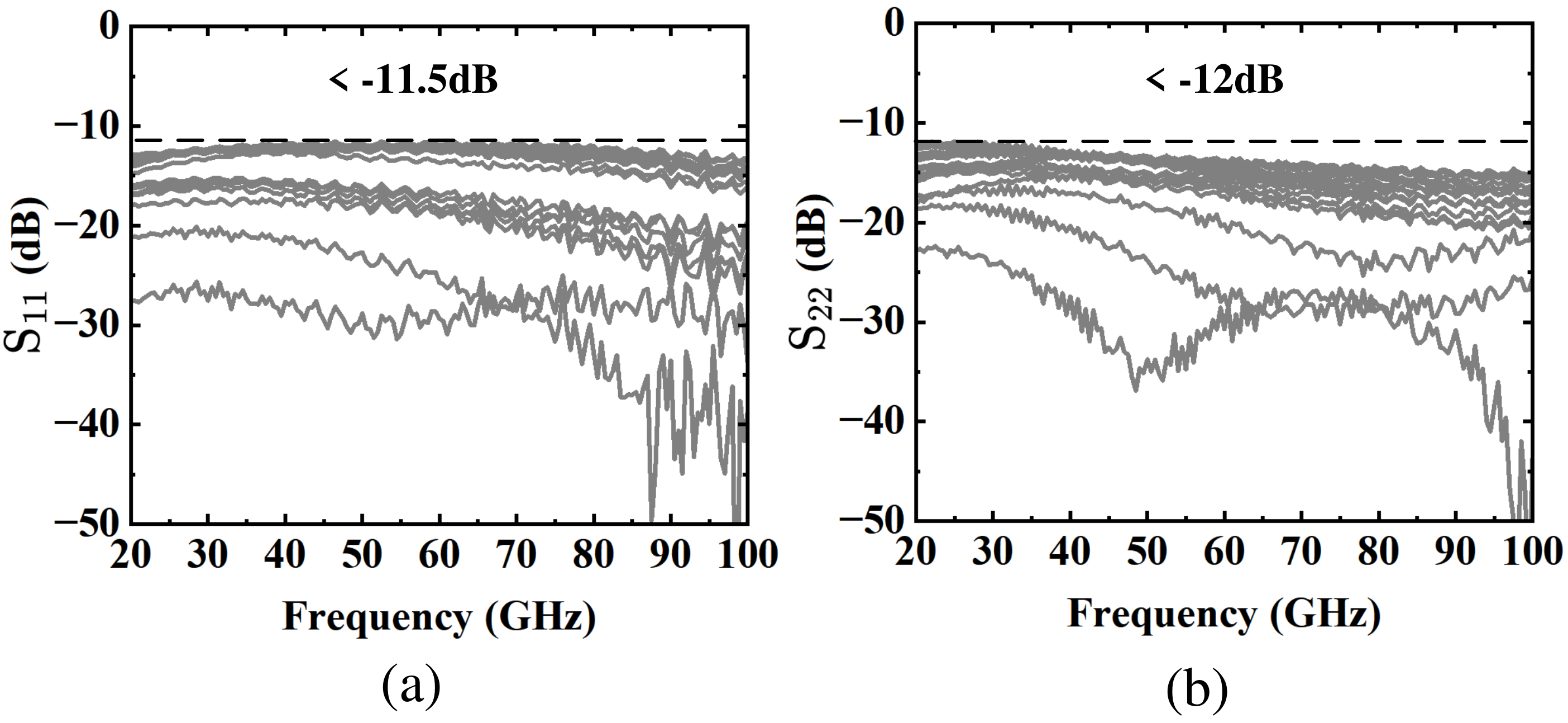}}
\caption{Measured (a) $\mathrm{S}_{11}$ and (b) $\mathrm{S}_{22}$ for 16 states.}
\label{RL}
\end{figure}

In Fig. \ref{Phase_RMS}, the phase responses for all states and RMS gain/phase errors are depicted. The RMS gain error and RMS phase error are lower than 0.15 dB and $1.6^\circ$, respectively, which indicates accurate amplitude tuning and low phase variation. The return losses at the input and output port remain below –11.5 dB, as shown in Fig. \ref{RL}, which indicates good matching of the attenuator.

The performance of the attenuator has been summarized and compared with state-of-the-art attenuator circuits in table I. 

\begin{table}[tbp]
\caption{COMPARISON WITH THE STATE$-$OF$-$THE$-$ART ATTENUATOR CIRCUITS}
\begin{center}
\renewcommand{\arraystretch}{1.5} 
\resizebox{0.48\textwidth}{!}{
\begin{tabular}{|c|c|c|c|c|c|}
\hline
    Reference & \textbf{This work} & \cite{ref1} & \cite{ref2} & \cite{ref3} & \cite{ref8} \\
\hline
\vspace{-0.1cm}
    \multirow{2}{*}{Technology} & \textbf{65nm} & 65nm & 65nm & 40nm & 40nm  \\
              & \textbf{CMOS} & CMOS & CMOS & CMOS & CMOS \\
\hline
    Frequency (GHz) & \textcolor{red}{\textbf{20--100}} & DC--50 & 10--50 & 6-15.3 & DC--87.8 \\
\hline
    ATT. Range/Step (dB) & \textbf{7.5/0.1} & 15.5/0.5 & 11/0.9 & 15.5/0.5 & 15.5/0.5 \\
\hline
    RMS ATT. Error (dB) & \textcolor{red}{\textbf{$<$0.15}} & $<$0.25 & $<$N/A & $<$0.3 & $<$0.5 \\
\hline
    RMS Phase Error (deg) & \textcolor{red}{\textbf{$<$1.6}} & $<$3.5 & $<$3 & $<$2.95 & $<$5.4 \\
\hline
    IL (dB)  & \textbf{1.6--3.8} & 1.5--5.9 & 0.8--4.5 & 1.7--3.2 & 1.4--5.3 \\
\hline
    DC Power (mW) & \textbf{0} & 0 & 0 & 18.4--29.6 & 0  \\
\hline
    RL (dB) & \textbf{$>$11.5} & $>$12 & $>$8 & $>$10 & $>$9  \\
\hline
    Core Area ($\text{mm}^2$) & \textcolor{red}{\textbf{0.015}} & 0.036 & 0.19 & 0.075 & 0.19 \\
\hline   
\end{tabular}
}
\label{tab1}
\end{center}
\end{table}

\section{Conclusion}
This work presents an attenuator that implemented in 
65nm CMOS technology with capacitive compensation to achieve ultra-broadband operation. In addition, the use of resistors implemented by metal lines, eliminating series transistors and isolated NMOS reduces the IL of the attenuator. TLs are inserted between attenuator units to achieve a good matching. The proposed attenuator can achieve a total attenuation of 7.5 dB with a broadband operation (20-100 GHz). The chip core area is $0.015\; \mathrm{mm^2}$. Its IL is only 1.6–3.8 dB over the operation band. The RMS gain error and RMS phase error of the attenuator are less than 0.15 dB and $1.6^\circ$, respectively.

\section*{Acknowledgment}
This work was supported by the National Natural Science Foundation of China under Grants 62371296, 62188102, and Okawa Foundation Research Grant.
\bibliographystyle{IEEEtran}
\bibliography{mybib}

\end{document}